\documentclass[12pt]{iopart}
\usepackage{graphicx}
\usepackage{color}

\begin{document}

\title[]{Nature of particles azimuthal anisotropy at low and high transverse momenta in ultrarelativistic A+A collisions}

\author{L V Bravina$^{1,2}$, G Kh Eyyubova$^2$, V L Korotkikh$^2$, I P Lokhtin$^2$, S V Petrushanko$^2$, A M Snigirev$^{2,3}$ and E E Zabrodin$^{2,1}$}

\address{$^1$ Department of Physics, University of Oslo, PB 1048 Blindern, N-0316 Oslo, Norway}
\address{$^2$ Skobeltsyn Institute of Nuclear Physics, Lomonosov Moscow State University, RU-119991 Moscow, Russia}
\address{$^3$ Bogoliubov Laboratory of Theoretical Physics, Joint Institute for Nuclear Research, RU-141980 Dubna, Russia }

\ead{eiiubova@lav01.sinp.msu.ru}
\vspace{12pt}

\begin{abstract}
LHC data on the correlations of the elliptic flow $v_2$ of particles at 
low and high transverse momenta $p_T$ from Pb+Pb collisions at 
center-of-mass energy per nucleon pair $\sqrt{s_{NN}} = 5.02$ TeV are 
analyzed in the framework of the HYDJET++ model. This model includes soft 
and hard components which allows to describe the region of both low and 
high transverse momenta. The origin of $v_2$ values in different $p_T$ 
regions is investigated at different centralities. It is shown that the 
experimentally observed correlations between $v_2$ at low and high $p_T$ 
in peripheral lead-lead collisions is due to correlation of particles in 
jets.
\end{abstract}

%
\vspace{2pc}

%
%
%
%

\section{Introduction}
The properties of hot and dense matter produced in heavy ion collisions 
are thoroughly explored during several decades, especially after the start 
of heavy-ion programs at the Relativistic Heavy Ion Collider (RHIC) at BNL
and the Large Hadron Collider (LHC) at CERN. The focus of the field is to 
investigate Quark Gluon Plasma (QGP) formation and its characteristics at 
different conditions. 
Numerous observables of final event serve to this purpose. Here we study 
one of the main observables in soft physics sector, namely the azimuthal 
momentum-space anisotropy of particles, and its manifestation in hard 
physics regime. The correlations between hard and soft contributions to 
azimuthal anisotropy of particles have attracted much attention, see 
\cite{Armesto}--\cite{Noronha}.

At relatively low transverse momenta, the azimuthal anisotropy occurs 
due to the anisotropic expansion of the compressed matter, since 
particles are emitted preferably in the direction of the largest pressure 
gradient \cite{ollitrault}. The anisotropy for particles with high 
transverse momenta is governed by the energy loss of hard partons 
traversing the hot and dense nuclear medium. Here more jet particles are
emitted in the direction of shortest path length \cite{gyulassy}. The 
sizable azimuthal anisotropy observed at RHIC energies was the main 
evidence for the nearly perfect liquid behavior of the created matter 
\cite{brahms_1,phobos_1,star_1,phenix_1}.
The parton energy loss in hot and dense medium, so-called jet quenching, 
can be investigated in experiment with different observables, such as 
nuclear modification factor of particle, jets suppression. The azimuthal 
particle asymmetry at high transverse momenta is yet another observable 
revealing information on jet quenching process. 
At LHC, the ATLAS \cite{atlas_flow} and CMS \cite{cms_flow} collaborations 
have performed measurements of the azimuthal anisotropy of charged 
particles produced in Pb+Pb collisions at $\sqrt{s_{NN}} = 5.02$~TeV up 
to $p_T = 60$ and 100 GeV/$c$, respectively.   

Anisotropic flow is quantitatively characterized by coefficients in the 
Fourier expansion of the azimuthal dependence of the invariant yield of 
particles in a form \cite{voloshin, poskanzer}:
 \begin{equation} 
   E\frac{d^3N}{dp^3} = \frac{1}{\pi}\frac{d^2N}{dp^2_Td\eta} \times 
   \left(1 + \sum\limits_{n = 1}^\infty 2v_n(p_T,\eta)
   \cos[n(\varphi -\Psi_{\rm {RP}})]\right),
 \end{equation}
where $\varphi$ is the azimuthal angle of particle, $\eta$ is the 
pseudorapidity and $\Psi_{\rm{RP}}$ is the reaction plane angle. The flow 
coefficients are: 
\begin{equation}
v_n =  \langle\langle \cos{ \left[ n(\varphi -\Psi_{\rm{RP}}) 
       \right]}\rangle\rangle,
\label{eq1}
 \end{equation}
where the averaging is performed over all particles in a single event and 
over all events. The initial reaction plane angle, which is defined by 
impact parameter vector, is not known, hence, the direct calculation of 
$v_2$ by (\ref{eq1}) is not possible.  
The measurement of collective flow coefficients aims to measure global 
azimuthal anisotropy, i.e., particle correlations with respect to global 
event geometry, but the methods of measuring are sensitive to local 
particle correlations (such as resonance decays, jets, fluctuations and 
so on). The most advanced experimental techniques, implementing gaps in 
pseudorapitidies between correlating particles, multiparticle cumulant 
methods and other, suggest suppression of local particle correlations. 

The aim of the present paper is to study the correlations between the
low-$p_T$ and high-$p_T$ regions of the elliptic flow of charged hadrons 
in Pb+Pb collisions at $\sqrt{s_{NN}} = 5.02$~TeV at different centralities.
These correlations were observed experimentally in \cite{atlas_flow} and
\cite{cms_flow}. The origin of the correlations is not understood yet,
mainly because of not too many models, such as EPOS \cite{epos},
QGSJet \cite{qgsjet} and HYDJET++ \cite{hydjet} (for recent developments,
see also \cite{ccnu} and \cite{mcgill}), which are describing the soft
and the hard processes simultaneously. 
Here we employ the phenomenological model HYDJET++ to explore the behavior 
of azimuthal particle anisotropies at different transverse momenta and its 
connection to common global event geometry, and also the feasibility to 
establish such connection by experimental techniques. 

\section{HYDJET++ model}
We use the HYDJET++ model \cite{hydjet} for nucleus-nucleus collision 
simulation which combines soft and hard physics as two components of 
resulting final heavy ion event. The soft physics is based on 
relativistic hydrodynamics and is represented by the thermal hadronic 
state generated on the chemical and thermal freeze-out hypersurfaces 
with preset freeze-out conditions. It includes the longitudinal, radial 
and elliptic flow effects and the decays of hadronic resonances. The code 
of it is based on the adapted version of the event generator FASTMC 
\cite{fastmc_1, fastmc_2}. The particle multiplicity distributions are 
Poissonians and their mean multiplicities are estimated within the 
effective thermal volume. The parameters for the soft part need to be 
tuned to describe experimental data. For instance, the elliptic flow of 
the produced particles is governed by the spatial and momentum anisotropy 
of the fireball. The parameter of the spatial anisotropy, $\epsilon_2(b)$,
regulates the elliptic profile of the final freeze-out hypersurface at a 
given impact parameter $b$. The momentum anisotropy parameter, $\delta(b)$, 
deals with the modulation of the flow velocity profile. Both parameters
are linked via the hydro-inspired parametrization \cite{fastmc_2}. 
Parameters responsible for spatial and momentum triangular anisotropy
are implemented in the model as well, thus giving rise to triangular 
and other odd harmonics flow.

The basis for hard physics of nucleus-nucleus collision is elementary QCD 
parton-parton scatterings as it is realized in PYTHIA \cite{pythia} with 
additional simulation of parton energy loss in a dense medium by PYQUEN 
model \cite{pyquen} with subsequent hadronization. Both collisional loss 
due to parton rescattering and gluon radiation loss are taken into account 
when propagating hard parton through a medium. The medium is treated as a 
boost-invariant longitudinally expanding medium at some temperature with 
the transverse asymmetric geometry given by impact parameter of a collision 
(initial elliptic shape is considered). The initial temperature of the 
medium for central collisions is one of the parameters of the model. The 
PYQUEN routine is used to generate a single hard nucleon-nucleon (NN) 
collision. To calculate the mean number of mini-jets produced in A+A
collision at a certain impact parameter $b$ one has to (i) determine the
number of binary NN collisions in the event and (ii) multiply it to the 
integral cross section of the hard process. Hardness
of the process depends solely on the minimum transverse momentum transfer, 
$p_{T}^{min}$. Therefore, the parameter $p_{T}^{min}$ is one of the major 
parameters which regulates the contributions of soft and hard particles to 
total multiplicity. If the transverse momentum of initial hard scattering
does not exceed $p_{T}^{min}$, the partons produced in the scattering are
excluded from the hard processes. The products of their hadronization are
then automatically added to spectrum of hadrons produced in soft processes.
\begin{figure}[htpb]
\centering
\includegraphics[scale=0.5]{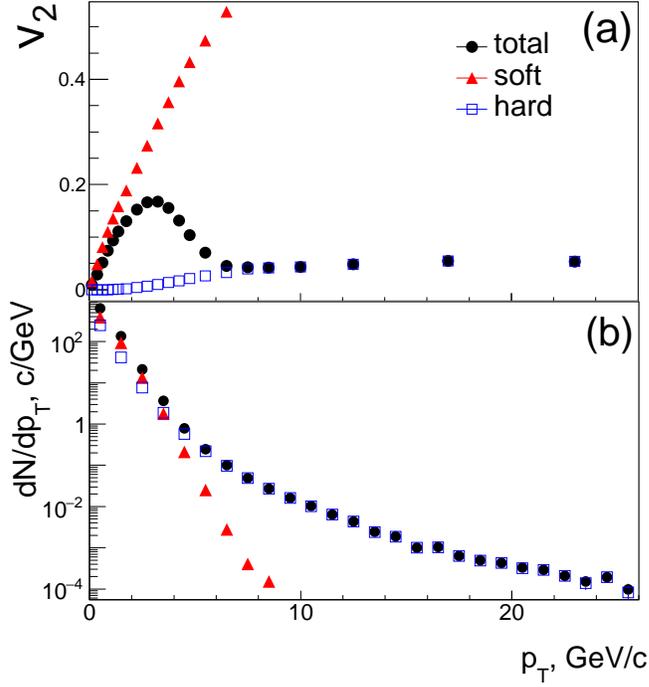}
\caption{Color online.
(a) The elliptic flow as a function of $p_T$, $v_2^{\rm{RP}}(p_T)$, 
for soft (triangles) and hard (squares) components, and resulting total flow 
(circles) of charged hadrons with $|\eta| < 2.4$ in 20--30\% centrality bin 
of Pb+Pb collisions calculated within HYDJET++ at $\sqrt{s_{NN}}= 5.02$ TeV.
(b) The same as (a) but for the transverse momentum spectra $dN/dp_T$ of 
charged hadrons for soft and hard components and also the total one.}
\label{fig1a_b}
\end{figure}
Figure \ref{fig1a_b}(a) demonstrates the contribution of the soft and 
hard components to the formation of elliptic flow in the model. 
Here the coefficients $v_2^{\rm{RP}}$ are calculated with respect to the 
reaction plane. 
Hydrodynamic nature of elliptic flow of the soft component (shown by red 
triangles) provides the increasing $v_2$ with rising $p_T$. For the hard 
component, non-zero elliptic flow arises at $p_T>10$~GeV/$c$ due to jet 
quenching effect in asymmetric medium. Figure \ref{fig1a_b}(b) shows the 
spectra $dN/dp_T$ of both components together with total spectrum. At low 
transverse momenta the multiplicity of soft particles dominates and one can 
see that the resulting flow in this region is determined mainly by the 
elliptic flow of soft component, while at high $p_T$ the hard component 
is dominant and determines total elliptic flow value. In the intermediate 
$p_T$ region the result is obtained by a simple superposition of two 
independent contributions.

HYDJET++ model has managed to describe many experimental features and 
phenomena measured in heavy ion collisions at RHIC and LHC energies. 
These signals include pseudorapidity and centrality dependence of the 
multiplicity of charged particles, their transverse momentum spectra, 
radii of $\pi\pi$ correlations in central Pb+Pb collisions \cite{malinina}, 
centrality and momentum dependencies of elliptic and higher-order 
components of the anisotropic flow \cite{hydjet_cqs}--\cite{hydjet_v3}, 
flow fluctuations \cite{hydjet_flow_fl}, angular dihadron correlations 
\cite{hydjet_ridge}, multiplicity correlations of charged particles in 
forward-backward hemispheres \cite{hydjet_fbcorr}, as well as effects
of jet quenching \cite{lokh1, lokh2} and, finally, production of heavy 
mesons \cite{hydjet_c1}--\cite{hydjet_c3}.

\section{Calculation of azimuthal anisotropy coefficients}
For the correct comparison with CMS data on $v_2$ of charged particles 
\cite{cms_flow} we apply the scalar product method (SP) and four-particle 
cumulant method \cite{cumulants}. We do not simulate the detector response, 
i.e., we use charged particles from the model, but apply the same 
$\eta$-cuts as in experimental data. The two- and four-particle correlations 
for the second order flow harmonic are defined as
\begin{equation} 
\eqalign{\langle\langle 2 \rangle\rangle = \langle \langle 
                         e^{i2(\varphi_1-\varphi_2)}\rangle\rangle, \cr
\langle\langle 4 \rangle \rangle = \langle \langle 
          e^{i2(\varphi_1+\varphi_2 -\varphi_3 -\varphi_4)}\rangle\rangle,} 
\label{eq:cor}
\end{equation} 
where double average means averaging over all particle combinations in an
event and over all events in a data sample.  
The estimator of the reference 4-particle cumulant, $c_2\{4\}$, is 
defined as
\begin{equation} 
c_2\{4\} = \langle\langle 4\rangle\rangle - \langle \langle 2\rangle\rangle^2
\end{equation} 
For the differential flow calculation one of the particle in (\ref{eq:cor}) 
is restricted to belong to a certain $p_T$ bin. We denote it by 
$\langle\langle 2^\prime \rangle\rangle$ and $\langle\langle 4^\prime 
\rangle\rangle$, respectively.   
The differential 4-particle cumulant reads
\begin{equation} 
d_2\{4\} = \langle\langle 4^\prime\rangle\rangle - 2\langle \langle 
2^\prime\rangle\rangle  \langle\langle 2\rangle\rangle.
\end{equation} 
The differential $v_2\{4\}(p_T)$ coefficient is derived as
\begin{equation} 
v_2\{4\}(p_T) = -d_2\{4\} (-c_2\{4\})^{-3/4}.
\end{equation}  
The cumulant calculations are based on $Q$-cumulant method \cite{q-cumulants}, 
where cumulants are expressed in terms of the corresponding $Q_n$ vectors.
The $v_2\{4\}(p_T)$ is calculated in midrapidity region $|\eta|<2.4$.  
Methods using many-particle correlations suppose to reflect global collective 
flow and to suppress local few-particle correlations. 
CMS measurement of $v_2$ at high $p_T$ \cite{cms_flow} has found similarity 
of results with 4-, 6-, 8-particle cumulants and also with SP method.   
The SP method also uses $Q_n$ vectors framework. The $Q_2$ vector for the 
second harmonic is defined as
\begin{equation} 
Q_2 = \sum\limits_{k = 1}^M \omega_k e^{i2\varphi_k},
\end{equation}   
where $M$ is the multiplicity of used particles and $\omega_k$ is a weight 
for a given particle $k$. Here we use SP method with three subevents, 
similarly to CMS calculation \cite{cms_flow}. 
Elliptic flow can be obtained as follows
\begin{equation}
v_2\{{\rm{SP}}\} = \langle Q_2 Q^*_{2A}\rangle / \sqrt \frac{\langle Q_{2A} 
Q^*_{2B}\rangle \langle Q_{2A} Q^*_{2C}\rangle}{ \langle Q_{2B} Q^*_{2C}
\rangle }.
\end{equation}   
The $Q_2$ vector for sub-events $A$ and $B$ is determined in $-5<\eta<-3$ 
and $3<\eta<5$ pseudorapidity regions, respectively, and for sub-event $C$ 
in $|\eta|<0.75$ region. The vectors $Q_{2A}$, $Q_{2B}$ and $Q_{2C}$ are 
calculated with weight $\omega$ equal to $p_T$ of particle. The $Q_2$ vector 
of particles of interest is calculated in $|\eta|<1$ pseudorapidity region 
with the unit weight. If the particle of interest comes with the positive 
$\eta$, then $Q_{2A}$ is calculated using the negative $\eta$ region, and 
vice versa. By its nature, the SP method resembles two-particle correlation 
methods, but using several sub-events with large $\eta$ gap suppresses 
few-particle correlations.   

\begin{figure}[htpb]
\centering
\includegraphics[scale=0.5]{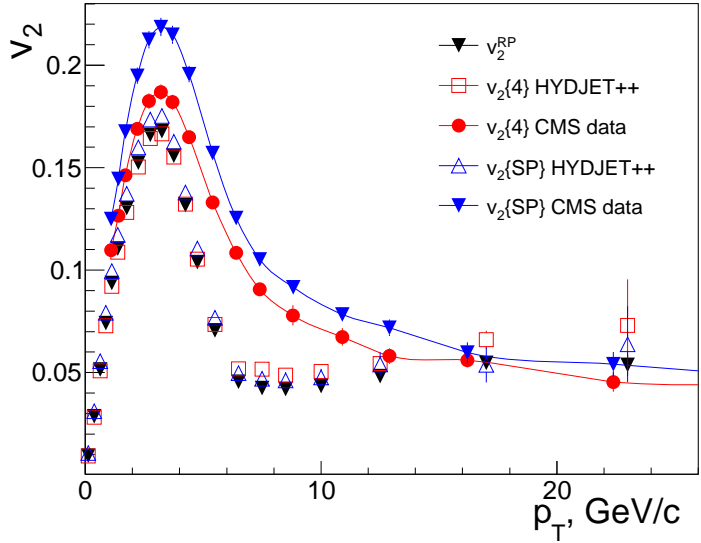} 
\caption{Color online.
The comparison of the model elliptic flow with CMS data 
\cite{cms_flow} for $v_2\{4\}(p_T)$ (data: circles; model: open squares) 
and $v_2\{\rm{SP}\}(p_T)$ (data: blue down triangles; model: open triangles) 
in Pb+Pb collisions at $\sqrt{s_{NN}}$ = 5.02~TeV for the centrality 20--30\%. 
The elliptic flow calculated with respect to reaction plane angle in the 
model, $v_2^{\rm{RP}}$, is also shown (black down triangles). Lines are drawn 
to guide the eye.}
\label{fig:2}
\end{figure}

In the model we also calculate elliptic flow $v_2^{\rm{RP}}$ with respect 
to reaction plane angle directly with (\ref{eq1}). $v_2^{\rm{RP}}$ is only 
associated with global event geometry. 
Figure \ref{fig:2} shows the elliptic flow $v_2\{4\}$, $v_2\{\rm{SP}\}$ and 
$v_2^{\rm{RP}}$ calculated from the HYDJET++ simulated events with 
centrality 20--30\% and compared to the CMS data \cite{cms_flow}. At 
relatively low transverse momenta, $p_T < 4$~GeV/$c$, the $v_2\{4\}(p_T)$ 
and $v_2\{\rm{SP}\}$ in HYDJET++ are similar to the generated original 
elliptic coefficient $v_2^{\rm{RP}}$. It is not surprising, because in this 
momentum region the bulk of the produced particles mainly correlates only 
with the reaction plane, whereas the non-flow correlations are quite small.
The model calculations are also in good agreement with the experimental 
results. Note, that in the data the difference between $v_2\{{\rm SP}\}(p_T)$ 
and $v_2\{4\}(p_T)$ is more pronounced. The model-generated $v_2\{\rm{SP}\}$ 
is closer to experimentally restored $v_2\{4\}(p_T)$ rather than to 
$v_2\{\rm{SP}\}$. The description of elliptic flow at other centralities can 
be found in \cite{hydjet_corr}. As was shown in \cite{hydjet_flow_fl}, a
better quantitative agreement of HYDJET++ results with the data can be 
reached if both anisotropy parameters, $\epsilon_2(b)$ and $\delta(b)$, are
treated as independent ones. In the region of high transverse momenta
with $p_T>10$ GeV/$c$ the model calculations fit to the data fairly well. 
In the region of intermediate transverse momenta with $4<p_T<10$ GeV/$c$, 
where we have transition between soft and hard physics, the model 
calculations are lower than the experimental data. However, this region is 
out of scope of our study and, therefore, it does not affect the results of 
the present work.  

\section{Elliptic flow correlation at low and high transverse momenta}
\begin{figure*}[htpb]
\centering
\includegraphics[scale=0.8]{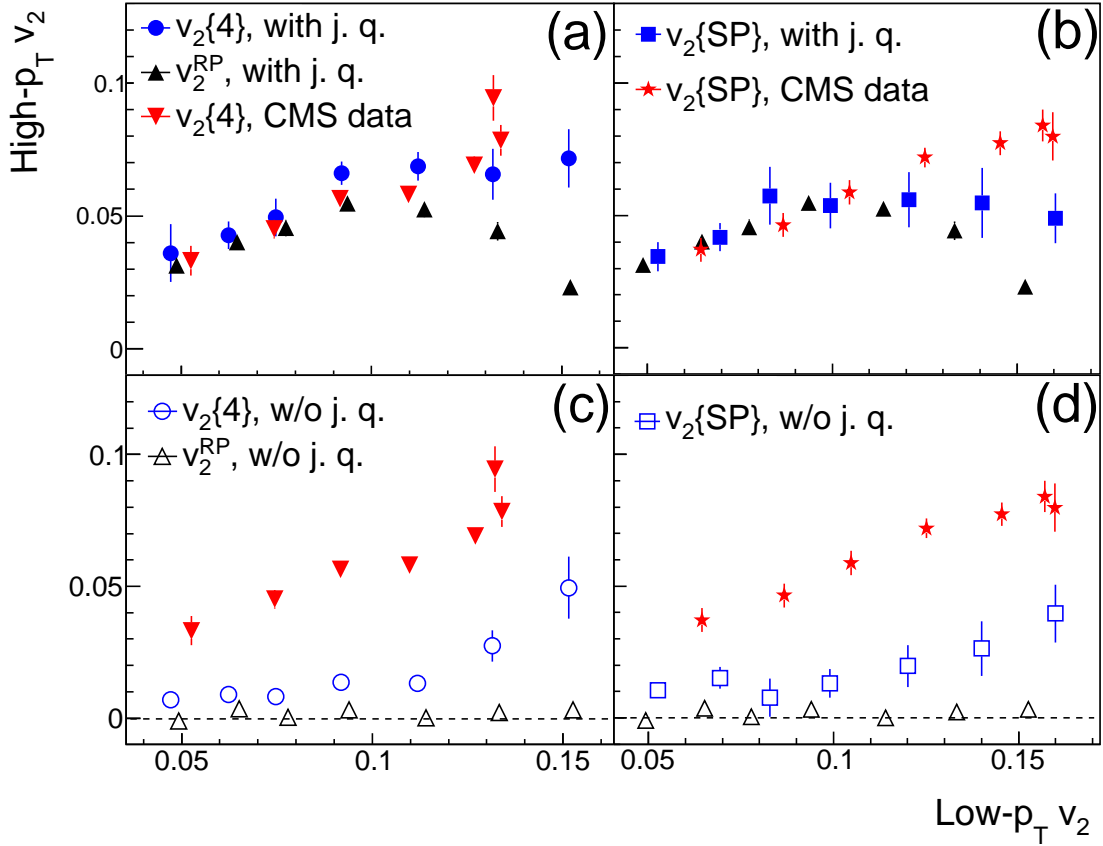} 
\caption{Color online.
The centrality dependence of the correlation between elliptic
flows of hadrons with low and high transverse momenta in Pb+Pb collisions 
at $\sqrt{s_{NN}}$ = 5.02~TeV. The results are shown for seven centrality 
bins 5--10\%, 10--15\%, 15--20\%, 20--30\%, 30--40\%, 40--50\%, and 50--60\%. 
Model calculations are presented for $v_2^{\rm{RP}}$ (black triangles), 
$v_2\{4\}$ (blue circles) and $v_2\{\rm{SP}\}$ (blue squares). CMS data for 
$v_2\{4\}$ (red down triangles) and $v_2\{\rm{SP}\}$ (red stars) are taken 
from \cite{cms_flow}. HYDJET++ calculations are performed also with jet 
quenching, j. q., (full symbols in (a) and (b)), and without jet quenching, 
(open symbols in (c) and (d)).}
\label{fig3a_d}
\end{figure*}

The lead-lead collisions at $\sqrt{s_{NN}}$ = 5.02~TeV were generated within
the HYDJET++ for the following centrality intervals: $\sigma/\sigma_{geo}$ = 
5--10\%, 10--15\%, 15--20\%, 20--30\%, 30--40\%, 40--50\% and 50--60\%.
The number of generated events for each centrality bin gradually increases
from $2 \times 10^6$ events for $\sigma/\sigma_{geo}$ = 5--10\% to 
$7\times 10^6$ events for $\sigma/\sigma_{geo}$ = 50--60\%. 
Figure \ref{fig3a_d} shows the correlations between the elliptic flow of
hadrons with transverse momenta $1.0 < p_T < 1.25$ GeV/$c$ and that of
hadrons with quite high transverse momenta, $14 < p_T < 20$ GeV/$c$. 
One can see that the HYDJET++ calculated flows, $v_2\{4\}(p_T)$ shown in
figure \ref{fig3a_d}(a) and $v_2\{\rm{SP}\}$ displayed in figure 
\ref{fig3a_d} (b), demonstrate the same the centrality dependence of the 
elliptic flow correlations as observed in the experiment.  
However, elliptic flow values calculated with respect to reaction plane angle, 
$v_2^{\rm{RP}}$, at low and high $p_T$ remain correlated only for semicentral 
collisions. For centralities larger than 30--40\% $v_2^{\rm{RP}}$ at high 
$p_T$ begins to die out. Elliptic flow $v_2^{\rm{RP}}$ at high $p_T$ appears 
solely due to jet quenching effect. To confirm this one can see the model 
simulations without jet quenching effect in figure \ref{fig3a_d}(c), where 
$v_2^{\rm{RP}}$ at high $p_T$ is consistent with zero. 
The jet quenching depends not only on path length of parton in a medium, but 
also on medium density and temperature. For peripheral collisions the 
asymmetry of path lengths increases but medium density decreases which leads 
to decreasing of $v_2^{\rm{RP}}$ at high $p_T$, see figure \ref{fig3a_d}(a). 
The coefficients $v_2\{4\}(p_T)$ and $v_2\{\rm{SP}\}$ have additional 
contribution to anisotropy rather than jet quenching effect, as can be seen 
from figure \ref{fig3a_d}(c), (d). Note that only jets contribute to the 
spectrum $dN/dp_T$ in this interval of transverse momentum in our approach 
and, therefore, the anisotropy can be caused by dijet topology.
Thus, we can distinguish at least two directions, namely, the jet axis and 
the reaction plane, relative to which the particles are correlated. The 
first one becomes more pronounced for peripheral collisions, which can be 
explained as the follows: in central collisions more jets are produced with 
random directions, for peripheral collisions the probability of dijet 
topology increases.  
The model simulations indicate that in collisions with centrality below 
30-–40\% azimuthal anisotropy occurs due to jet quenching, while in more 
peripheral collisions, jet particle correlations make a significant 
contribution to the fourth order cumulant and scalar product methods.

\section {Conclusions}
The analysis of azimuthal anisotropy coefficients at low and high transverse
momenta has been performed for Pb+Pb collisions generated within
two-component model HYDJET++ at center-of-mass energy $\sqrt{s_{NN}}$ = 
5.02~TeV per nucleon pair.
Both phenomena, responsible for the origin of azimuthal anisotropy at low 
and high $p_T$, namely hydrodynamic expansion of the created matter and jet 
quenching, are related and sensitive to initial anisotropy of overlap region 
of nuclei. We have found the correlation of elliptic flow values measured 
with respect to reaction plane angle, $v_2^{\rm{RP}}$, at low and high $p_T$ 
in central and mid central collisions (up to 40\%). For more peripheral 
collisions the $v_2^{\rm{RP}}$ at high $p_T$ decreases.   
Nevertheless, for coefficients measured by cumulant and scalar product 
methods additional source of azimuthal anisotropy emerges for peripheral 
collisions , i.e., anisotropy connected to dijet topology. 
If the centrality of the collisions is 30--40\% or less, the four-cumulant 
and the scalar product methods are sensitive mainly to the azimuthal 
anisotropy of initial overlapping region. As the collisions become more 
peripheral, the azimuthal anisotropy begins to be determined primarily by 
the correlation of particles inside the jets. Combination of these two 
origins of anisotropy tends to reproduce experimentally observed 
correlations of the elliptic flow values at different transverse momenta 
in all centralities.
 
\ack
We would like to thank A.I. Demyanov, L.V. Malinina and A.V. Belyaev for 
fruitful discussions. This work was supported in parts by Russian Foundation 
for Basic Research (RFBR) under Grants No. 18-02-00155, No. 18-02-40084 and 
No. 18-02-40085. LVB and EEZ acknowledge support of the Norwegian Research 
Council (NFR) under grant No. 255253/F50 “CERN Heavy Ion Theory.”

\end{document}